\documentstyle[epsf,11pt]{article}

\newtheorem{theorem}{Theorem}

\newenvironment{proof}%
 {\par\noindent{\underline{Proof} \quad}}{\hfill$\Box$\bigskip}
 {\par\noindent{\underline{Proof} of the theorem\quad}}{\hfill$\Box$\bigskip}
 {\par\smallskip\noindent{\underline{{\it Remark}} \quad}}{\par\smallskip}
 {\par\smallskip\noindent{\underline{{\it Fact}} \quad}}{\par\smallskip}
\newenvironment{example}%
 {\par\smallskip\noindent{\underline{{\it Example}} \quad}}{\par\smallskip}
 {\par\smallskip\noindent{{\it Assumotion} \quad}}{\par\smallskip}
 {\par\smallskip\noindent{{\it Condition} \quad}}{\par\smallskip}

\newcommand{\ol}{\overline}

\newcommand{\tr}{{\rm tr}}
\newcommand{\Tr}{{\rm Tr}}
\newcommand{\R}{{\bf R}}
\newcommand{\C}{{\bf C}}

\newcommand{\CR}{{\rm CR}}

\newcommand{\rgl}{\rangle}
\newcommand{\lgl}{\langle}

\begin{document}

\begin{center}

\vspace*{2mm}
{\Large Berry's phase in view of quantum estimation theory,\\
   and\\
its intrinsic relation with the complex structure}\\
\vspace{10mm}
Keiji Matsumoto
\footnote{
Department of Mathematical Engineering and Information Physics \\
University of Tokyo, Bunkyo-ku,Tokyo 113, Japan }
\begin{abstract}
In this paper, it is pointed out that the Berry's phase 
is a good index of degree of no-commutativity of
 the quantum statistical model.
Intrinsic relations between the `complex structure' of
the Hilbert space and Berry's phase is also discussed.
\end{abstract}

{\it Keywords: Berry's phase
    quantum estimation theory, 
     attainable Cramer-Rao type bound, 
    complex structure, antiunitary operator}
        
\end{center}

\section{Introduction}
Berry's phase,
convinced by many experiments,
is naturally interpreted as a curvature of natural connection introduced
on the line bundle over the space of pure states
\cite{AA}\cite{Berry}\cite{Shapere}\cite{Simon}.

In this paper, it is shown that the Berry's phase 
is a good index of non-commutativity of the quantum statistical model,
and that Berry's phase has intrinsic relation
with the `complex structure' of the Hilbert space.

The paper is organized as follows. 
Section \ref{sec:preliminaries} is review of quantum estimation theory
and Berry's phase. 
In section \ref{sec: metric}, 
sections \ref{sec:berryest}-\ref{sec:berryglb},
relations between Berry's phase and quantum estimation theory are discussed.
In section \ref{sec:d} and sections \ref{sec:antiunitary}-{sec:timereversal},
relations between Berry's phase and the `complex structure' is studied.

\section{Preliminaries}
\label{sec:preliminaries}
\subsection{Quantum measurement theory, the unbiased estimator}
We denote by ${\cal P}_1({\cal H})$
the space of  density operators of pure states 
in a  separable Hilbert space ${\cal H}$.
${\cal P}_1({\cal H})$ is often simply denoted by  ${\cal P}$,
${\cal P}_1 $.

Let $\Omega$ be a space of all possible outcomes of 
an experiment,
and $\sigma(\Omega)$ be a $\sigma$- field in $\Omega$.
When the density operator of the system is $\rho$,
the probability  that the data $\omega \in \Omega$
lies in $B\in\sigma(\Omega)$ writes 
\begin{eqnarray}
{\rm Pr}\{ \omega \in B|\rho \} =\tr \rho M(B),
\label{eqn:pdm}
\end{eqnarray}
by use of the map  $M$ from $\sigma(\Omega)$
to  nonnegative Hermitian operator which satisfies
\begin{eqnarray}
&&M(\phi)=O, M(\Omega)=I,\nonumber\\
&&M(\bigcup_{i=1}^{\infty} B_i),
=\sum_{i=1}^{\infty}M(B_i)
\;\;(B_i\cap B_j=\phi,i\neq j),
\label{eqn:maxiom}
\end{eqnarray}
so that $(\ref{eqn:pdm})$ define a probability measure
(see Ref.{\rm \cite{Helstrom:1976}}, p.53 
and Ref.{\rm \cite{Holevo:1982}}, p.50).
We call the map $M$ the {\it measurement}, 
because there always exist an physical experiment corresponds to the map $M$
which satisfies $(\ref{eqn:maxiom})$
\cite{Steinspring:1955}\cite{Ozawa:1984}.

The quantum estimation theory deals with
identification of the density operator of the given physical system
from the data obtained by the experiment.
For simplicity, we usually assume that
the density operator is a member of a {\it model}, or a manifold of
${\cal M}
=\{\rho(\theta)|\theta\in \Theta \subset {\bf R}^m\}\subset{\cal P}$,
and that
the finite dimensional parameter $\theta$ is to be estimated statistically.
For example, ${\cal M}$ is 
the set of spin states with given wave function part and unknown spin part.
In this paper, we restrict ourselves to the {\it pure state model} case,
where any member of the model is pure state,
\begin{eqnarray}
\rho(\theta)&=&\pi(|\phi(\theta)\rgl)\nonumber\\
			&\equiv&|\phi(\theta)\rgl\lgl\phi(\theta)|\nonumber.
\end{eqnarray}

To estimate the parameter, we perform the experiment
and obtain the data $\omega$ by which we calculate
an estimator $\hat\theta$ by {\it estimator } $\hat\theta(\omega)$.
A pair $(\Omega,\,\sigma(\Omega),\, M,\, \hat\theta)$ 
 of an estimator $\hat\theta(*)$,
a measurement $M$,
a space $\Omega$ of data, a $\sigma$-field $\sigma(\Omega)$
is also called an {\it estimator}. 
The expectation of $f(\omega)$ with respect to the probability distribution 
$(\ref{eqn:pdm})$ is denoted by $E_\theta[f(\omega)|M]$.

The estimator $(\Omega,\,\sigma(\Omega),\, M,\, \hat\theta)$ 
 is said to be  {\it unbiased} if 
\begin{eqnarray}
E_\theta[\hat\theta(\omega)|M]=\theta
\label{eqn:unbiased}
\end{eqnarray}
holds for all $\theta\in\Theta$.
If $(\ref{eqn:unbiased})$ and
\begin{eqnarray}
\partial_i E_\theta[\theta^j(\omega)|M]=\delta^j_i\:(i,j=1,...,m),
\nonumber
\end{eqnarray}
where $\partial_i$ stands for $\partial/\partial \theta^i$,
hold at a 
particular  $\theta$, $(\Omega,\,\sigma(\Omega),\, M,\, \hat\theta)$
 is called {\it locally unbiased} at $\theta$. 

As a measure of the error of the locally unbiased estimator,
we use the weighed sum of covariance matrix,
\begin{eqnarray}
\sum_{i=1}^m g_i [V_{\theta}[\hat\theta(\omega)|M]]_{ii},
\label{eqn:sgv}
\end{eqnarray}
where $V_{\theta}[\hat\theta(\omega)|M]$ is the covariance matrix of 
the random variable $\hat\theta(\omega)$ which obeys 
the probability distribution $(\ref{eqn:pdm})$, and
$g_i\:9i=1,...,m)$ are non-negative real numbers,
which corresponds to the `cost' caused by the wrong estimation
of the true value of the $i$-th component of the parameter.
More generally, 
\begin{eqnarray}
\Tr G V_{\theta}[\hat\theta(\omega)|M],
\end{eqnarray}
where $G$ is a real symmetric nonnegative matrix,
is also used as a measure of the error, and its infimum
\begin{eqnarray}
\CR(\theta,G,{\cal M})
=\inf\{\Tr G V_{\theta}[\hat\theta(\omega)|M]\: |\:
        \mbox{$(\Omega,\,\sigma(\Omega),\, M,\,\hat\theta)$ is 
                locally unbiased at $\theta$}\}
\end{eqnarray}
is called the {\it attainable Crmaer-Rao (CR) type bound},
in analogy with the Cramer-Rao inequality in the classical estimation theory
(throughout the paper, the term `classical estimation theory'
means the estimation theory of the family of probability distributions). 
The real symmetric nonnegative matrix $G$ is called {\it weight matrix}.
For $\CR(\theta,\, G,\, {\cal M})$, 
abbreviated notations such as $\CR(\theta,\, G)$ $\CR(G)$ are often used.

\subsection{SLD CR inequality, the attainable CR type bound}
We have the following {\it SLD CR inequality},
which is proved for the exact state model 
by Helstrom \cite{Helstrom:1967}\cite{Helstrom:1976},
and is proved for the pure state model by Fujiwara and Nagaoka
\cite{FujiwaraNagaoka:1995}:
\begin{eqnarray}
\CR(\theta,\, G) \geq \Tr G \left(J^S(\theta)\right)^{-1}.
\label{eqn:mpCR}
\end{eqnarray}
Here  $J^S(\theta)$, called {\it SLD Fisher information matrix}, 
is defined by
\begin{eqnarray}
J^S(\theta)\equiv \left[{\rm Re}\lgl l_i(\theta)|l_j(\theta)\rgl\right].
\end{eqnarray}
where  $|\l_i(\theta)\rgl\:(i=1,...,m)$ are defined below.

The {\it horizontal lift}
$|l_X(\theta)\rgl$ of a tangent vector
$X\in {\cal T}_{\rho(\theta)}({\cal M})$ to $|\phi(\theta)\rgl$,
is an element of ${\cal H}$ which satisfies
\begin{eqnarray}
X\rho(\theta)=\pi_*(|l_X\rgl)\equiv
\frac{1}{2}(|l_X\rgl\lgl\phi(\theta)|+|\phi(\theta)\rgl\lgl l_X|),
\label{eqn:sld:lift}
\end{eqnarray}
and
\begin{eqnarray}
\lgl l_X|\phi(\theta)\rgl=0.
\label{eqn:sld:horizontal}
\end{eqnarray}
Here, 
$X$ in the left hand side $(\ref{eqn:sld:lift})$ of is to be understood as
a differential operator.
$|l_i(\theta)\rgl$ is defined to be a horizontal lift of 
$\partial_i\in {\cal T}_{\rho(\theta)}({\cal M})$ .

The uniqueness of the horizontal lift is proved easily.
As for the existence, given a manifold 
${\cal N}=\{ |\phi(\theta)\rgl\,\: \, \theta\in\Theta\}$
 in ${\cal H}$ such that
\begin{eqnarray}
{\cal M}&=&\pi({\cal N})\nonumber\\
&\equiv&\{\rho(\theta)\: |\: \rho(\theta)=\pi(\theta),\,\theta\in\Theta\},
\end{eqnarray}
the horizontal lift  explicitly writes
\begin{eqnarray}
|l_X(\theta)\rgl= 2X|\phi(\theta)\rgl -2\rho(\theta)X|\phi(\theta)\rgl.
\end{eqnarray}

The inequality $(\ref{eqn:mpCR})$ is of special interest,
because of the following theorems.
\begin{theorem} (Fujiwara and Nagaoka \cite{FujiwaraNagaoka:1995})
If the model is one dimensional,  
the equality in $(\ref{eqn:mpCR})$ establishes.
\label{theorem:1pCR}
\end{theorem}
\begin{theorem}\cite{Keiji:1997}
If the weight matrix is in the form of 
\begin{eqnarray}
G=diag(0,..,0, g_i, 0,..., 0),
\end{eqnarray}
the equality in $(\ref{eqn:mpCR})$ establishes.
\label{theorem:evievi}
\end{theorem}

However, the bound is not always attainable\cite{Keiji:1996}.
\begin{theorem}
The  SLD CR inequality is 
attainable for a strictly positive definite weight matrix,
iff 
\begin{eqnarray}
{\rm Im}\lgl l_i(\theta)|l_j(\theta)\rgl=0
\label{eqn:imlilj=0}
\end{eqnarray}
for any $i,j$.
\label{th:imlilj=0}
\end{theorem}
The model is said to be {\it locally quasi-classical} at $\theta$
iff $(\ref{eqn:imlilj=0})$  holds true at $\theta$.

In general,
\begin{eqnarray}
&&\inf \left\{\left. \sum_{i=1}^m g_i [V_{\theta}[M]]_{ii}\, \right|\, 
          \mbox{ $M$ is locally unbiased at $\theta$}\right\}\nonumber\\
&\geq&
\sum_{i=1}^m \inf \left\{\left. g_i [V_{\theta}[M]]_{ii}\, \right|\, 
          \mbox{ $M$ is locally unbiased at $\theta$}\right\}\nonumber\\
&=&
\sum_i  g_i \left[J^{S-1}(\theta)\right]^{ii},
\nonumber
\end{eqnarray}
holds true, where the last equality comes from theorem $\ref{theorem:evievi}$,
and if the model is not quasi-classical and the weight matrix is
strictly positive, the equality in first  inequality
does not establish.
The gap between the both sides of the inequality 
can be considered to be caused by the non-commutativity of 
quantum theory.

\subsection{Berry's phase}
In this section, we review the geometrical theory of Berry's phase.

Let us denote by $\tilde{\cal H}$ the totality  of state vectors,
or member of ${\cal H}$ with unit length.
Because the two  state vectors
correspond to the same state
iff  they  differ with each other only in their phase factor,
it is natural to consider $\tilde{\cal H}$
as a principal fiber bundle with the base space ${\cal P}_1$
and the structure group $U(1)$.

A {\it horizontal lift} $\hat{C}=\{|\phi(t)\rgl\,|\, 0\leq t \leq 1\}$ of 
the curve $C=\{\rho(t)\,|\, 0\leq t \leq 1\}$ in ${\cal P}_1$
is defined to be a curve in $\tilde{\cal H}$
which satisfies
$\rho(t)=\pi(|\phi(t)\rgl)$  and
\begin{eqnarray}
| l_{d/dt} (t)\rgl=2\frac{d}{dt}|\phi(t)\rgl.
\end{eqnarray}
The connection introduced by this horizontal lift is
called {\it Pancharatnam connection}.
Then, the Berry's phase $\gamma(C)$ 
for the curve $C=\{\rho(t)\,|\, 0\leq t \leq 1\}$
is defined by
\begin{eqnarray}
|\phi(1)\rgl=e^{i\gamma(C)}|\tilde{\phi}(1)\rgl,
\end{eqnarray}
where $|\tilde{\phi}(1)\rgl$ satisfies 
$\pi(|\tilde{\phi}(1)\rgl)=\rho(1)$ and
\begin{eqnarray}
{\rm Im}\lgl \phi(0)|\tilde{\phi}(1)\rgl = 0.
\end{eqnarray}
The model is said to be {\it parallel} iff
Berry's phase for any curve in the model vanishes.

The Berry's phase for the infinitesimal loop 
\begin{eqnarray}
\begin{array}{ccc}
(\theta^1,..., \theta^i,..., \theta^j+d\theta^j,..., \theta^m)
&\leftarrow&
(\theta^1,..., \theta^i+d\theta^i,..., \theta^j+d\theta^j,...., \theta^m)
\\
\downarrow& &\uparrow
\\
\theta=(\theta^1,..., \theta^i,..., \theta^j,...., \theta^m)
&\rightarrow&
(\theta^1,..., \theta^i+d\theta^i,..., \theta^j,...., \theta^m)
\end{array}
\label{loop}
\end{eqnarray}
is calculated up to the second order of $d\theta$
as 
\begin{eqnarray}
\frac{1}{2}\tilde{J}_{ij}  d\theta^i d\theta^j + o(d\theta)^2,
\nonumber
\end{eqnarray}
where $\tilde{J}_{ij}$ is equal to
${\rm Im}\lgl l_i|l_j \rgl$ .

Mathematically,
\begin{eqnarray}
\frac{1}{2}\sum _{i,j}\tilde{J}_{ij}d\theta^i d\theta^j
\label{eqn:berrycurv}
\end{eqnarray}
is an representation of  the curvature form of Pancharatnam connection.

\section{SLD and Fubini-Study metric}
\label{sec: metric}
It should be noted that the SLD FIsher information matrix is
deeply concerned with this fiber bundle structure.
Given a curve $C=\{\rho(t)\}$ in ${\cal P}_1$,
let us consider the minimization
\begin{eqnarray}
\min\left\{\left.
  \lgl\dot{\psi}(t)|\dot{\psi}(t)\rgl\}\right|
  \pi(|\psi(t)\rgl )=\rho(t)\right\},
\end{eqnarray}
which corresponds to `the shortest distance' between infenitesimally distant
fibers $\pi^{-1}(\rho(t))$ and $\pi^{-1}(\rho(t+dt))$.
It is pointed out that
the minimum is achived by the horizontal lift
\cite{Abe}, and 

For any curve $\hat{C}'=\{|\phi(t)\rgl\,|\, 0\leq t \leq 1\}$
such that  $C=\pi(\hat{C}')$, we can prove the inequality
\begin{eqnarray}
V_t[M\,|\, \hat{t}]\geq \frac{1}{4 \lgl\dot{\phi}'(t)|\dot{\phi}'(t)\rgl}\,,
\end{eqnarray}
almost in the same way as the SLD CR inequality,
where $(\Omega,\,\sigma(\Omega),\, M, \hat{t})$ is 
a locally unbiased estimator of the parameter $t$, 
and dot `` $\dot{ }$ '' stands for the differentiation with respect to $t$.
Most strict inequality of this type  is obtained by
the minimization of the denominator of the right hand side of the inequality.
It is already known that the minimum is given by 
horizontal lift $\hat{C}$. Given this fact, it is easily understood
that the minimum is equal to the SLD Fisher information matrix. 

Therefore, SLD Fisher information $J^S_t(t)$ of the parameter $t$
is proportional 
to `the minimum distance' between infinitesimally distant two fibers 
$\pi^{-1}(\rho(t))$ and $\pi^{-1}(\rho(t+dt))$,
and the minimization corresponds to the search of the best possible bound.

Remember that 
the inverse of the Fisher information matrix is
the attainable lower bound of the covariance matrix of the unbiased estimator
when the model is 1-dimensional. This fact implies
that if SLD Fisher information is smaller,
it is harder to distinguish $\rho(t)$ and $\rho(t+dt)$.
We can also say that
the closer 
	two fibers
			$\pi^{-1}(\rho(t))$ and $\pi^{-1}(\rho(t+dt))$
				are,
the harder
	it is
	  to distinguish $\rho(t)$ 
		 from $\rho(t+dt)$.

In the following, 
we define inner product $\lgl * , * \rgl_{\theta}$ 
in ${\cal T}_{\theta}({\cal M})$ by
\begin{eqnarray}
\lgl \partial_i , \partial_j \rgl_{\theta}
=\left[ J^S(\theta) \right]_{ij},
\label{eqn:defmetric}
\end{eqnarray}
because  this metric ({\it SLD metric}, hereafter) seems to be
estimation-theoretically  and geometrically natural.
Notice the unique existence of the horizontal lift assures
that the equality $(\ref{eqn:defmetric})$
certainly defines a metric.

\section{${\bf D}$-transform }
\label{sec:d}
We define a linear transform ${\bf D}_{\theta}$
in ${\cal T}_{\theta}({\cal M})$ by
\begin{eqnarray}
\lgl \partial_i,{\bf D}_{\theta}(\partial_j)\rgl_{\theta}
=\tilde{J}_{ij}(\theta).
\label{eqn:defd}
\end{eqnarray}
${\bf D}_{\theta}$ is called {\it D-transform}.
The non-zero eigenvalues of ${\bf D}_{\theta}$ are denoted by
$\pm i\beta_j(\theta)$,
where $\beta_j(\theta)$ is positive real number,
and $j$ runs from $1$ to the half of the rank of D-transform,
and $\beta_j(\theta)s$ are sorted so that $\beta_1\geq\beta_2\geq ...$.

When $\dim {\cal M}=2$, 
\begin{eqnarray}
\beta_1(\theta)=\frac{\tilde{J}_{12}(\theta)}{\sqrt{\det J^S(\theta)}}, 
\end{eqnarray}
left hand side of which is Berry's 
along the curve which encloses unit area,
where the unit of area is naturally induced by the SLD metric.

It is worthy of remarking that the ${\bf D}_{\theta}$ is
a manifestation of 
the natural complex structure of the Hilbert space ${\cal H}$.
Actually, $D$-transform is obtained by the following procedure.

First,multiply the imaginary unit $i$ to 
$|l_X\rgl$. 
Since
$i|l_X\rgl$  
is not a horizontal lift of any element of 
${\cal T}_{\rho}({\cal M})$ generally,
we project $i|l_X\rgl)$ to $span_{\R}\{|l_1\rgl,...,|l_m\rgl\}$
with respect to the metrc ${\rm Re}\lgl * | * \rgl$.
${\bf D}X\in {\cal T}_{\rho}({\cal M})$ is defined to be
a tangent vector whose horizontal lift is identical to
the product of the projection.

Defining $D$-transform in this way,
the curevature form of Pancharatnam connection is defined by
the equation $(\ref{eqn:defd})$.
Therefore, Berry's phase is a manifestation of the complex structure
of the Hilbert space. 

\begin{eqnarray}
\begin{array}{ccc}
        &\mbox{multiplication of}\:\: i& 	
\\
|l_X\rgl  \in span_{\bf R}\{|l_1\rgl,...,|l_m\rgl\}&--\longrightarrow
&i|l_X\rgl  
\\
\uparrow& &\,\downarrow 
\\
\mbox{taking horizontal lift}& &\mbox{project w.r.t.\, ${\rm Re}\lgl*|*\rgl$,} 
\\
 & &\mbox{reverse image of the horizontal lift}
\\
\uparrow& &\downarrow 
\\
X\in{\cal T}_{\rho}({\cal M})&--\longrightarrow
& {\bf D}X\in{\cal T}_{\rho}({\cal M})
\\
  &{\bf D}&
\end{array}
\nonumber
\end{eqnarray}

\section{Berry's phase and local non-commutativity}
\label{sec:berryest}
In this section, it is pointed out that 
Pancharatnam curvature
is good index of non-commutativity of two components of the parameter.
As is already mentioned, 
the difference between 
$\CR(G)$ and $\Tr G J^{S-1}$
 is an manifestation of the non-commutativity of the quantum mechanics.
The author conjectures the difference
increases as $\beta_j$s increase, because of the following reasons.

First, in the 2-dimensional pure state model, 
we have the following theorem \cite{Keiji:1996}.
\begin{theorem}
Let ${\cal M}$ and ${\cal M}'$ be a 2-dimensional pure state model.
Then,
if $\beta_1(\theta', {\cal M}')\geq \beta_1(\theta, {\cal M})$
and $J^S(\theta', {\cal M}')=J^S(\theta, {\cal M})$,
then for any weight matrix $G$,
\begin{eqnarray}
\CR(G, \theta', {\cal M}')\geq \CR(G, \theta, {\cal M}).
\label{eqn:cr>cr}
\end{eqnarray}
\end{theorem}

Second, by virtue of theorem $\ref{th:imlilj=0}$,
if Pancharatnam curvature vanishes,
the gap between both sides of 
the inequality $(\ref{eqn:cr>js})$ vanishes.

Third, 
as for  general multi-dimensional pure state models,
we have \cite{Keiji:1996},
\begin{eqnarray}
&&\CR\left(\theta,J^{S}(\theta),{\cal M}\right)\nonumber\\
&=&\left\{\Tr
     {\rm Re}
       \sqrt{I_m+i J^{S-1/2}(\theta)\tilde{J}(\theta)J^{S-1/2}(\theta)}
     \right\}^{-2}\nonumber\\
&=&\sum_j
\frac{4}{1+\sqrt{1-|\beta_j(\theta)|^2}}.
\label{eqn:trjsv}
\end{eqnarray}

Notice that the transform of the parameter,
\begin{eqnarray}
\theta^i \rightarrow \theta^{'i}=a_i\theta^i,
\nonumber
\end{eqnarray}
and the change of the weight, 
\begin{eqnarray}
g_i \rightarrow g'_{i}=a_i^2 g_i,
\nonumber
\end{eqnarray}
effects the attainable CR type bound in the exactly the same manner.
Therefore, if every component are to be equally weighed in the sum,
the weight matrix must be the one which `normalize' the `length'
of each component.
When the model is parameterized so that  $J^S$ is diagonal,
the weight matrix $J^S$ gives one of good normalizations,
because $[J^{S-1}]^{ii}$ is minimum variance of 
locally unbiased estimator of the submodel of ${\cal M}$
such that all the components of the parameter other than
$\theta^1$ are fixed to some constant.

However, in the general case, the estimation theoretical meaning 
of $\CR(J^S)$
is hard to verify. Still, this value is geometrical in the sense 
that it remains invariant under
any transform of the coordinate in the model ${\cal M}$.
Considering the non-commutativity of the model 
should be invariant under the coordinate transform,
$\CR(J^S)$ can be one of good measures of non-commutativity.

\section{Berry's phase and global  non-commutativity}
\label{sec:berryglb}
Even if  the model is locally quasi-classical at any $\theta\in\Theta$,
the best locally unbiased estimator 
$(\Omega,\,\sigma(\Omega),\, M,\,\hat\theta)$ is  dependent on true value of 
the parameter $\theta$. 
A locally quasi-classical model is said to be {\it quasi-classical}
when the measurement $M$ in the best pair 
$(\Omega,\,\sigma(\Omega),\, M,\,\hat\theta)$ 
is independent of true value of the parameter,
for,  in this case, 
using  the globally best measurement $M$,
the adaptation of the calculation of the estimate from the data,
which is the problem of the classical estimation theory,
is only left to be done to find the globally best pair
 $(\Omega,\,\sigma(\Omega),\, M,\,\hat\theta)$.

In the faithful model case, or the case where the model is consisted of
the strictly positive density operator, 
it is pointed out the model quasi-classical
iff Uhlmann's RPF(relative phase factor) 
vanishes for any curve in the model{{Keiji:1997a}.
For Uhlmann's RPF is nothing but a generalization of Berry's phase 
to the non-pure model, it is of interest to study if the analogical fact
holds true for the pure state model.

For simplicity,
we say that the manifold ${\cal N}$ in ${\cal P}_1$ 
is a {\it horizontal lift} of the model ${\cal M}$ if
\begin{eqnarray}
&&\pi({\cal N})={\cal M},\\
&&\forall |\phi(\theta)\rgl\in {\cal N},\,
2\frac{\partial}{\partial \theta^i}|\phi(\theta)\rgl
=|\l_i(\theta)\rgl,
\nonumber
\end{eqnarray} 
holds true.
The horizontal lift ${\cal N}$ exists iff ${\cal M}$ 
is locally quasi-classical at any point.
The model is said to be {\it quasi-parallel}
iff  Berry's for any curve is $0$ or $\pi$.

\begin{theorem}
If  the model ${\cal M}$ is quasi-parallel, that model is  
quasi- classical.
\label{th:p-plgl1}
\end{theorem}
\begin{proof}
First, apply  Schmidt's orthonormalization to
the horizontal lift ${\cal N}$ of ${\cal M}$,
to obtain the orthonormal basis ${\bf B}=\{ |e_i\rgl\,|\, i=1, 2, ...\}$
such that ${\cal N}$ is a subset of the real span of ${\bf B}$.
We immerse Hilbert space ${\cal H}$ into $L^2(\R,\,\C)$
as 
\begin{eqnarray}
|\phi(\theta)\rgl
=\sum_i a_i|e_i\rgl\mapsto \sum_i a_i(\theta)\psi_i(x),
\nonumber
\end{eqnarray}
where $\{\ psi_i(x)\,|\, i=1, 2, ...\}$ is an orthonormal basis
in $L^2(\R,\,\C)$. Then, letting $E(dx)$ be a projection valued measure
which is obtained as a spectral decomposition of the position operator,
the pair $(\R,\, \sigma(\R),\, E,\,\hat\theta)$ is 
one of the best estimators.
This assertion is easily proved by calculating 
the Fisher information matrix of the family,
\begin{eqnarray}
\left\{ p(x\, , \theta)=\sum_i (a_i(\theta))^2 |\psi_i(x)|^2
\right\}
\end{eqnarray}
of probability distributions.
\end{proof}

\begin{example}
Let $\psi(x)$ is a wave function, or an element of 
$L^2(\R^d, \C)$, then, the model 
\begin{eqnarray}
&&{\cal M}_x(\psi(x))=\pi({\cal N}_x)\nonumber\\
&&{\cal N}_x(\psi(x))=\left\{ |\phi(\theta)\rgl\: \left|\: 
		|\phi(\theta)\rgl=\psi(x-\theta),
		\theta\in\R^d \right.\right\},
\nonumber
\end{eqnarray}
is said to be a {\it (d-dimensional) position shifted model},
and the {\it (d-dimensional) momentum shifted model} ${\cal M}_p(\psi(x))$
and the {\it (d-dimensional)  position-momentum shifted model} 
 ${\cal M}_{x,p}(\psi(x))$
are defined almost in the same way.

If $\psi(x)$ takes real value only and is symmetric about orgin
(, for esample, the eigenstates of the harmonic oscilator),
then the position shifted model ${\cal M}_x(\psi(x))$ and
the momentum shifted model  ${\cal M}_p(\psi(x))$ are quasi-parallel,
and, therefore, are quasi-classical.
\end{example}
\begin{example}
Denote by  $J_z$ the $z$-component of the spin operator ${\bf J}$,
by $|m\rgl\,(m=-S,-S+1,...,S)$ the $m$-th eigenstate of $J_z$,
where $S$ is the total spin,
and define the model ${\cal M}_{J_z}(|\psi\rgl)$ by
\begin{eqnarray}
&&{\cal M}_{J_z}(\psi\rgl)=\pi({\cal N}_{J_z}(|\psi\rgl))
\nonumber\\
&&{\cal N}_{J_z}(|\psi\rgl)=
\{|\phi(\theta)\rgl\: |\: 
    e^{-i\frac{\theta}{\hbar}J_z}|\phi(\theta)\rgl,\, \theta\in\R\}.
\nonumber
\end{eqnarray} 
If and only if  $|\psi\rgl$ satisfies
\begin{eqnarray}
|\lgl m|\psi\rgl |=|\lgl -m|\psi\rgl |\:\:\: (m=-S, -S+1,..., S),
\nonumber
\end{eqnarray}
the model ${\cal M}_{J_z}(|\psi\rgl)$ is quasi-parallel and
quasi-classical.
\end{example}
\begin{example}
The Riemanian Geodesic with respect to the metric tensor $J^S(\theta)$
is quasi-parallel and quasi-classical.
\end{example}

The converse of the theorem is, however, not true, because
the following counter-examples exist.

\begin{example}
We consider the model ${\cal M}_x$ which is defined by 
\begin{eqnarray} 
&&{\cal M}_x=\pi({\cal N}_x)\nonumber\\
&&{\cal N}_x=\left\{ |\phi(\theta)\rgl\: \left|\: 
		|\phi(\theta)\rgl=
		c\mbox{\it const.}\times(x-\theta)^2 e^{-(x-\theta)^2+ig(x-\theta)},\:
		\theta\in\R \right.\right\},
\nonumber
\end{eqnarray}
where $g$ the function such that
\begin{eqnarray}
g(x)=\left\{
\begin{array}{cc}
0 & (x\geq 0),\\
\alpha & (x< 0).
\end{array}
\right.
\nonumber
\end{eqnarray}

Then, as easily checked,
${\cal N}_x$ is a horizontal lift of the model ${\cal M}_x$,
and 
$\lgl\phi(\theta)|\phi(\theta')\rgl$ is not real 
unless $\alpha=n\pi\,(n=0,1,...)$.
However, SLD CR bound is uniformly attained by
the measurement obtained 
by the spectral decomposition $E(dx)=|x\rgl\lgl x|dx$ 
of the position operator,where $|x_0\rgl=\delta(x-x_0)$.
as is  checked by
comparing SLD Fisher information of the model ${\cal M}_x$
and the classical Fisher information of 
the probability distribution family
\begin{eqnarray}
\left\{ p(x|\theta)\,\left|\, 
	p(x|\theta)=|\lgl\phi(\theta)|x\rgl|^2,\;
	\theta\in\R\right.\right\}.
\nonumber
\end{eqnarray}

Note that $|\phi(\theta)\rgl$ is 
an eigenstate of the Hamiltonian 
\begin{eqnarray}
H(\theta)=-\frac{\hbar^2}{2m}\frac{d^2}{dx^2}
	+\frac{\hbar^2}{m}\left( 2(x-\theta)^2+\frac{1}{(x-\theta)^2}\right),
\nonumber
\end{eqnarray}
whose potential  has two wells
with infinite height of wall between them.
\end{example}

\begin{example}
Let ${\cal H}$ be $L^2([0,\,2\pi],\,\C)$,
and  define a one parameter model ${\cal M}$ such that,
\begin{eqnarray}
&&{\cal M}=\pi({\cal N})\nonumber\\
&&{\cal N}=\left\{|\phi(\theta)\rgl\: \left|\: 
				|\phi(\theta)\rgl=\mbox{\it const.}\times(2-\cos\omega)\, 
	e^{i\alpha(f(\omega-\theta)+\theta)},\:(0\leq\omega, \theta< 2\pi) 
	\right.\right\},
\nonumber\\
\label{eqn:ring}
\end{eqnarray}
where $\alpha$ is a real number and
$f$ the function defined by
\begin{eqnarray}
f(\omega-\theta)=\left\{
\begin{array}{cc}
\omega-\theta & (\omega-\theta\geq 0)\\
\omega+2\pi-\theta &(\omega-\theta < 0)
\end{array}
\right. .
\nonumber
\end{eqnarray}
Physically, $(\ref{eqn:ring})$ is an eigenstate of the Hamiltonian $H$
such that,
\begin{eqnarray}
H(\theta)=-\frac{\hbar^2}{2m}\left(\frac{d}{d\omega}-i\alpha\right)^2
+\frac{A-B\cos(\omega-\theta)}{2-\cos(\omega-\theta)},
\nonumber
\end{eqnarray}
which characterize the dynamics of an electron
confined to the one-dimensional ring which encircles magnetic flax
$\Phi=2\pi\alpha c/e$, where $m$ is the mass of the electron,
$-e$ the charge of the electron, $c$ the velocity of light.
$A$ and $B$ are the appropriately chosen constant.

It is easily checked that ${\cal N}$ is  a  horizontal lift of the model
${\cal M}$,
and that 
the model ${\cal M}$ is not parallel unless $\alpha= n\pi\,(n=0,1,...)$.
However, consider
the projection valued measure $E_{\omega}$ such that
\begin{eqnarray}
E_{\omega}(d\omega)=|\omega\rgl\lgl\omega| d\omega,
\nonumber
\end{eqnarray}
where $|\omega_0\rgl=\delta(\omega-\omega_0)$.
Then, it is easily checked that
the classical Fisher information of 
the probability distribution family
\begin{eqnarray}
\left\{ p(\omega|\theta)\,\left|\,  
p(\omega|\theta)=|\lgl \phi(\theta)|\omega\rgl |^2,\,
0\leq\omega,\theta< 2\pi\right.\right\}
\nonumber
\end{eqnarray}
is equal to the SLD Fisher information of ${\cal M}$.
\end{example}

\section{The antiunitary operator and Time reversal symmetry}
\label{sec:antiunitary}

\subsection{The antiunitary operator}
As is pointed out in the section $\ref{sec:d}$, 
Berry's phase seems to have
 some intrinsic relation with the `complex structure'.
In this section, we study this point using the antiunitary operator.

The  transformation $A$ 
\begin{eqnarray}
|\tilde{a}\rgl=A|a\rgl,\:\:\:  |\tilde{b}\rgl=A|b\rgl
\nonumber
\end{eqnarray}
is said to be {\it antiunitary} iff
\begin{eqnarray}
\lgl \tilde{a} |\tilde{b}\rgl&=&\ol{\lgl a| b \rgl},\nonumber\\
A(\alpha|a\rgl+\beta|b\rgl)&=&\ol{\alpha}A|a\rgl+\ol{\beta}A|b\rgl,
\nonumber
\end{eqnarray}
where $\ol{z}$ means complex conjugate of $z$ (see Ref \cite{Sakurai} p.266).

\begin{theorem}
The model is quasi-parallel iff the horizontal lift of the model is 
invariant by some antiunitary operator. 
\end{theorem}

\begin{proof}
Suppose that any member of the manifold 
${\cal N}=\{|\phi \rgl\}$ in $\tilde{\cal H}$ 
is invariant by the antiunitary operator $A$,
and let $|\tilde{\phi}\rgl=A|\phi\rgl,\,|\tilde{\phi}'\rgl=A|\phi'\rgl$.
Then, we have
\begin{eqnarray}
\lgl \phi |\phi'\rgl=\lgl \tilde{\phi'}|\tilde{\phi}\rgl=\lgl \phi'|\phi\rgl
\in\R.
\nonumber
\end{eqnarray}
Conversely, if $\lgl \phi|\phi'\rgl$ is real for any 
$|\phi\rgl,\,|\phi'\rgl\in{\cal N}$,
by Schmidt's orthonormalization, we can obtain 
the orthonormal basis ${\sf B}=\{|i\rgl\,|\, i=1,2, ... ,d\}$
such that  ${\cal N}$ is subset of the real span of ${\sf B}$,
which means any member of ${\cal N}$ is invariant by 
the antiunitary operator $K_{\sf B}$, which is defined by,
\begin{eqnarray}
K_{\sf B}\sum_i\alpha_i |i\rgl= \sum_i \ol{\alpha}_i|i\rgl.
\nonumber
\end{eqnarray}
\end{proof}


\subsection{Time reversal symmetry}
As an example of the antiunitary operator,
we discuss {\it time reversal operator} 
(see Ref.\cite{Sakurai}, pp. 266-282).
The time reversal operator $T$ is an antiunitary operator
in $L^2(\R^3,\,\C)$ which transforms 
the wave function $\psi(x)\in L^2(\R^3,\,\C)$ as:
\begin{eqnarray}
T \psi(x)=\ol{\psi(x)}=K_{\{|{\bf x}\rgl\}}\psi(x).
\nonumber
\end{eqnarray}
The term `time reversal' came from the fact
that if  $\psi({\bf x}, t)$ is a solution of
the Sch\"odinger equation
\begin{eqnarray}
i\hbar\frac{\partial\psi}{\partial t}=
\left(-\frac{\hbar^2}{2m}\nabla^2+V\right)\psi,
\nonumber
\end{eqnarray}
then $\ol{\psi({\bf x}, -t)}$ is also its solution.

The operator $T$ is sometimes called {\it motion reversal operator},
since it transforms the momentum eigenstate 
$e^{i{\bf p}\cdot{\bf x}/\hbar}$ corresponding to eigenvalue ${\bf p}$
to the eigenstate  $e^{-i{\bf p}\cdot{\bf x}/\hbar}$
corresponding to eigenvalue $-{\bf p}$.

Define the {\it position shifted model} by
\begin{eqnarray}
{\cal M}_{\bf x}=
\{\rho(\theta)\,|\, 
\rho(\theta)=\pi(\psi({\bf x}-{\bf x}_0)\,),\,{\bf x}_0 \in\R^3\},
\nonumber
\end{eqnarray}
and suppose that any member of 
the horizontal lift ${\cal N}_{\bf x}$
of the model ${\cal M}_{\bf x}$
has time reversal symmetry.
Then, since time reversal operator $T$ is antiunitary,
the model  ${\cal M}_{\bf x}$
is quasi-classical in the wider sense.
The spectral decomposition of the position operator
gives optimal measurement.

Now, we discuss the generalization of time reversal operator.
In this paper, the antiunitary transform
\begin{eqnarray}
T_{\alpha}\, : \, e^{i{\bf p}\cdot{\bf x}/\hbar}
\rightarrow
e^{i\alpha({\bf p})}\,e^{-i{\bf p}\cdot{\bf x}/\hbar}
\nonumber
\end{eqnarray}
is also called
motion reversal operator, or time reversal operator.

If the wave function $\psi({\bf x})$ 
is invariant by the time reversal operator $T_{\alpha}$,
\begin{eqnarray}
\int_{\R^3} 
\psi({\bf x}-{\bf x}_0)\,\ol{\psi({\bf x}-{\bf x}'_0)}\,
d{\bf x}
\in\R
\label{eqn:psipsi}
\end{eqnarray}
holds true for any ${\bf x}_0,\, {\bf x}'_0$,
which means  that the position shifted model ${\cal M}_{\bf x}$
is quasi-parallel and, therefore, is quasi-classical.

Conversely, if $(\ref{eqn:psipsi})$ holds true,
Fourier transform of $(\ref{eqn:psipsi})$ leads to
\begin{eqnarray}
|\Psi({\bf p})|^2=|\Psi(-{\bf p})|^2,
\nonumber
\end{eqnarray}
where
\begin{eqnarray} 
\Psi({\bf p})=
\frac{1}{\sqrt{2\pi}}
\int\psi({\bf x})e^{-i{\bf p}\cdot{\bf x}/\hbar} d{\bf x}.
\nonumber
\end{eqnarray}
Therefore,
the wave function $\psi({\bf x})$ is transformed  to itself
by the time reversal operator $T_{\alpha}$
such that
\begin{eqnarray}
T_{\alpha}\, : \, e^{i{\bf p}\cdot{\bf x}/\hbar}
\rightarrow
e^{i(\beta({\bf p})+\beta(-{\bf p})\,)}\,e^{-i{\bf p}\cdot{\bf x}/\hbar},
\nonumber
\end{eqnarray}
where
\begin{eqnarray} 
e^{i\beta({\bf p})}=\frac{\Psi({\bf p})}{|\Psi({\bf p})|}.
\nonumber
\end{eqnarray}

\begin{theorem}
A position shifted model is quasi-parallel
if and only if  there exists 
the time reversal operator which transforms
the wave function $\psi({\bf x})$ to itself.
\end{theorem}

\section*{Acknowledgment}
The author is grateful to Dr. A. Fujiwara, 
for introducing the quantum estimation theory to the author,
and for inspiring discussions.
The author is also thankful to Dr. H. Nagaoka for constructive discussions.

\end{document}